\documentclass[11pt]{iopart} 
\usepackage{amsmathred}
\usepackage{amsfonts}
\usepackage{graphicx}
\usepackage{epsfig}
\usepackage{subfigure}
\usepackage{bbm}
\usepackage{psfrag}
\usepackage{pstool}
\usepackage{amsbsy}
\newcommand{\ba}{\begin{align}}
\newcommand{\ea}{\end{align}}
\newcommand{\be}{\begin{equation}}
\newcommand{\ee}{\end{equation}}
\newcommand{\bea}{\begin{eqnarray}}
\newcommand{\eea}{\end{eqnarray}}

\begin{document}
\title{\sf \bfseries Phonon creation by gravitational waves}
\author{\sf \bfseries Carlos Sab{\'\i}n $^{1}$, David Edward Bruschi $^{2}$, Mehdi Ahmadi $^{1}$, Ivette Fuentes $^{1}$}
\address{
$^{1}$ School of Mathematical Sciences,
University of Nottingham,
University Park,
Nottingham NG7 2RD,
United Kingdom\\ 
$^{2}$Racah Institute of Physics and Quantum Information Science, 
The Hebrew University of Jerusalem,
91904, Givat Ram, Jerusalem,
Israel}
\ead{c.sabin.les@gmail.com}
\begin{abstract}
We show that gravitational waves create phonons in a Bose-Einstein condensate (BEC). A traveling spacetime distortion produces particle creation resonances that correspond to the dynamical Casimir effect in a BEC phononic field contained in a cavity-type trap. We propose to use this effect to detect gravitational waves.  The amplitude of the wave can be estimated applying recently developed relativistic quantum metrology techniques. We provide the optimal precision bound on the estimation of the wave's amplitude.  Finally, we show that the parameter regime required to detect  gravitational waves with this technique could be, in principle, within experimental reach in a medium-term timescale.
\end{abstract}

\maketitle
\section {Introduction.} Einstein's theory of general relativity \cite{rindlerrelativity} predicts the existence of gravitational waves \cite{reviewgravwaves}. Gravitational waves are perturbations of the spacetime generated by accelerated mass distributions. The theory predicts that the amplitude of gravitational waves is extremely small and thus, finding experimental evidence of their existence is a difficult task. Indeed the quest for the detection of these spacetime distortions \cite{gravwavesdetectors} has been one of the biggest enterprises of modern science and the focus of a great amount of work, both in theory and experiment. 
 \begin{figure}[t]
\includegraphics[width=\linewidth]{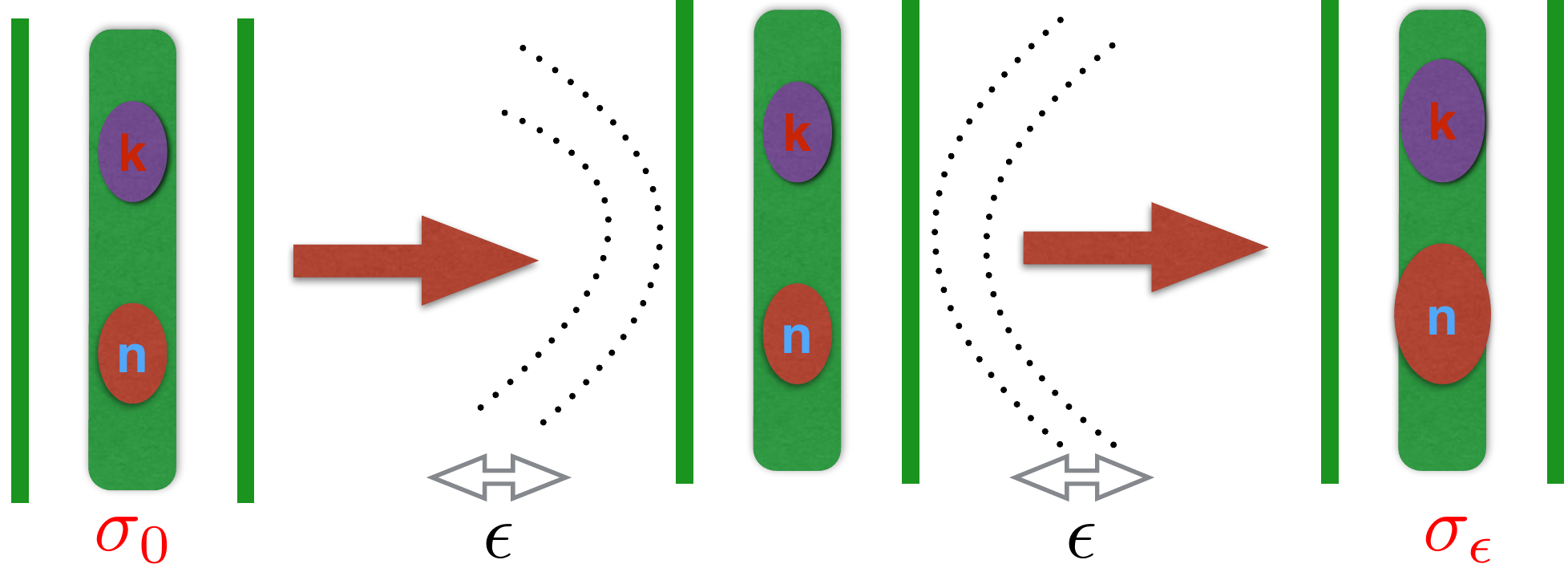}
\caption{Sketch of the setup: A BEC in a box-like potential that acts as a cavity for the phononic excitations. Two phononic modes $k$ and $n$ are initially prepared in a two-mode squeezed state represented by the initial covariance matrix $\sigma_0$. A gravitational wave of amplitude $\epsilon$ transforms the state producing excitations. The new state $\sigma_{\epsilon}$ depends on $\epsilon$. Measurements on the modes can be used to estimate the amplitude of the spacetime distortion.}\label{fig:fig1}
\end{figure}

In this paper we show that small spacetime distortions produce phononic excitations in a BEC.  We propose a scheme that exploits this effect to detect gravitational waves.  Phononic excitations of a trapped BEC satisfy a Klein-Gordon equation on a curved background metric. The metric has two terms \cite{matt, liberati,salelites}, one corresponding to the real spacetime metric and a second term, corresponding to the analogue gravity metric, which depends on BEC parameters such as velocity flows and energy density. In the field of analogue gravity, the real spacetime metric is considered to be flat while the BEC parameters are modified by the experimentalist in order to mimic spacetime dynamics. In this way, sonic black holes and expanding universes are simulated in a BEC. Interestingly, the effects of the real spacetime metric on the phononic field have been ignored.  In this paper, we show that changes in the real spacetime metric produce phononic excitations that can be in principle detected. This generalises the work presented in reference \cite{rqm}  to the curved background case. 
In particular we consider the case of gravitational waves such as those created by supernovas and gamma-ray bursts. We propose to use this effect, which is both a quantum and a relativistic effect, to design a novel gravitational wave detector. By applying recently developed techniques in relativistic quantum metrology \cite{rqm,rqm2}, we show that the regime of functionality of the phononic gravitational wave detector is within experimental reach.

Quantum metrology \cite{advances} exploits quantum properties to improve the performance of measurement technologies. Indeed, quantum metrology promises to provide useful techniques for gravitational wave detection \cite{reviewquantumgrav} and are currently being used in one of the most ambitious programs of gravitational-wave astronomy, the Laser Interferometry Gravitational-Wave Observatory (LIGO) \cite{gravwaveastronomy}.  LIGO is a laser interferometer that attempts to detect gravitational waves by means of the changes they produce in the optical paths of the laser in the interferometer's arms. Small path changes should produce phase shifts that can be measured at the output of the interferometer. Since the gravitational waves are very small, the interferometer arms must be 4 km long and the mirrors' weight is 10 kg. LIGO's sensitivity can be enhanced employing laser quantum states, such as squeezed states. By this method, the noise can be in principle reduced to shot-noise quantum limit \cite{quantumgravobs}.  Unfortunately, the regimes at which these techniques would provide important advantages, have not been reached yet in LIGO. Sources of error such as the thermal noise of the test masses, are still challenges to overcome before one can hope to reach the quantum regime. An alternative method that has been proposed to detect gravitational waves using quantum states involves an atom interferometer  \cite{sagas, hogan}. In an atom interferometer, the wavefunction of a BEC is split and recombined by means of laser pulses, giving rise to a phase shift proportional to the acceleration induced spacetime distortion. Interestingly, although both laser and atom interferometer schemes operate in the overlap of quantum mechanics and relativity, they do not consider relativistic quantum field theoretical effects, such as particle creation \cite{moore, casimirwilson} or mode-mixing \cite{bsgates}. The schemes mentioned above use non-relativistic quantum mechanics on one hand, and general relativity on the other. However, these theories are known to be incompatible.

Quantum field theory in curved spacetime allows one to properly incorporate quantum and relativistic effects at low energies. The energy regime in which the theory is applicable includes the natural energy regime of gravitational waves that are produced by, for example, supernovas. Interestingly, the application of metrology techniques to quantum field theory in curved spacetime remains practically unexplored. Only recently, a framework for relativistic quantum metrology has been developed applying metrology techniques to estimate parameters of quantum fields that undergo relativistic transformations \cite{rqm, rqm2, aspachs,downes}. The techniques can be applied to estimate spacetime parameters, proper times, gravitational field strenghts and accelerations, among other quantities of great interest to science and technology.  Using this framework, it was shown that the non-uniform acceleration of a BEC cavity trap produces phononic excitations that can be detected with cutting-edge technology. This effect occurs in a flat spacetime metric with moving boundary conditions, i.e. corresponding to the dynamical Casimir effect. Phonon creation provides the basis of a quantum accelerometer that exploits relativistic effects to improve the state of the art in quantum accelerometers \cite{rqm}. 

In this paper, we apply relativistic quantum metrology techniques to calculate the optimal precision bound achieved for detecting gravitational waves using the phononic creation effect we described above.  We characterise the transformation that a gravitational wave generates on the phononic excitations in a BEC (see Fig. \ref{fig:fig1}). The final state depends on the amplitude of the gravitational wave. We compute the quantum Fisher information associated to this state and -via the quantum Cramer-Rao bound- we obtain the fundamental bound to the error in the measurement of  the spacetime ripple. We find the experimental regime of parameters for which this bound is low enough to enable the detection of a gravitational wave.

\section{Modelling the gravitational wave spacetime}  
The metric of a gravitational wave spacetime is commonly modelled by a small perturbation $h_{\mu\nu}$ to the flat Minkowski metric $\eta_{\mu\nu}$, i.e. \cite{reviewgravwaves}, \begin{equation}\label{eq:gmunu}
g_{\mu\nu}=\eta_{\mu\nu}+h_{\mu\nu}
\end{equation}
where 
\begin{equation}\label{eq:metricplane}
\eta_{\mu\nu}= \begin{pmatrix}-c^2&0&0&0\\0&1&0&0\\0&0&1&0\\0&0&0&1\end{pmatrix}.
\end{equation}
and $c$ is the speed of light in the vacuum. We consider Minkowski coordinates $(t,x,y,z)$. In the transverse traceless (TT) gauge \cite{reviewgravwaves}, the perturbation corresponding to a gravitational wave moving in the z-direction can be written as,
\begin{equation}
h_{\mu\nu}= \begin{pmatrix}0&0&0&0\\0&h_{+}(t)&h_{\times}(t)&0\\0&h_{\times}(t)&-h_{+}(t)&0\\0&0&0&0\end{pmatrix},
\end{equation}
where $h_{+}(t)$, $h_{\times}(t)$ correspond to time-dependent perturbations in two different polarisations. Later on we will restrict the analysis to 1-dimensional fields, where the line element takes a simple form,
\begin{equation}\label{eq:lineelement}
ds^2=-c^2\,dt^2+(1+h_{+}(t))\,dx^2.
\end{equation}
The null geodesics of the above line element yield the speed of propagation of photons in the spacetime, which is $c(t)=\frac{dx}{dt}=\simeq c(1-\frac{1}{2}\,h_{+}(t))$ up to the first order in h.\\

\section{Bose-Einstein condensates on a  curved and flat spacetime} 
We are interested in describing a BEC in this spacetime and exploring the possibility that the gravitational waves produce observable effects in the system. The system is extremely small, we will consider the BEC  to be 1 $\operatorname{\mu\,m}$ long. Thus, it is natural to think that the effects are too small to be observable. Surprisingly, amplification effects produced by the slow propagation of excitations on the BEC make the effects, in principle, observable. To show this we describe the BEC on a general spacetime metric following references \cite{matt, liberati,salelites}.  In the superfluid regime, a BEC  is described by a mean field classical background $\Psi$ plus quantum fluctuations $\hat\Pi$. These fluctuations, for length scales larger than the so-called healing length, behave like a phononic quantum field on a curved metric.  Indeed, 
in a homogenous condensate, the field obeys a massless \footnote{We consider only the massless modes, since the massive modes -which require an energy comparable to the chemical potential of the condensate- are not excited by the gravitational wave} Klein-Gordon equation $\Box\hat\Pi=0$ where the d' Alembertian operator $\Box=1/\sqrt{-\mathfrak{g}}\,\partial_{a}(\sqrt{-\mathfrak{g}}\mathfrak{g}^{ab}\partial_{b})$ depends on an effective spacetime metric $\mathfrak{g}_{ab}$ -with determinant $\mathfrak{g}$- given by  \cite{matt,liberati,salelites}
\begin{equation}
\mathfrak{g}_{ab}=\left(\frac{n^2_0\,c_s^{-1}}{\rho_0+p_0}\right)\left[g_{ab}+\left(1-\frac{c_s^2}{c^2}\right)V_aV_b\right].
\end{equation}
The effective metric is a function of the real spacetime metric $g_{ab}$ (that in general may be curved) and   
background mean field properties of the BEC such as the number density $n_0$, the energy density $ \rho_0$, the pressure $p_0$ and the speed of sound $c_s=c\sqrt{\partial p/\partial\rho}$.  Here $p$ is the total pressure, $\rho$ the total density and $V_a$ is the 4-velocity flow on the BEC. In the field of analogue gravity, the real spacetime metric is considered to be flat and, therefore, it's effects are neglected. Analogue spacetimes are simulated through the artificial manipulation of what we call the analogue gravity metric , 
\begin{equation}
\mathfrak{G}_{ab}=\left(\frac{n^2_0\,c_s^{-1}}{\rho_0+p_0}\right)\left[\left(1-\frac{c_s^2}{c^2}\right)V_aV_b\right].
\end{equation}
Experimentalists change the background parameters of the analogue metric $\mathfrak{G}_{ab}$ to simulate sonic black holes or expanding universes \cite{analoguereview2011}. Here we are interested solely on the effects of the real spacetime metric. To ensure this, we consider that in the comoving frame $V_a=(c,0,0,0)$ and obtain,
\begin{equation}\label{eq:effmetric}
\mathfrak{g}_{ab}= \left(\frac{n^2_0\,c_s^{-1}}{\rho_0+p_0}\right) \left[g_{ab}+ \begin{pmatrix}(c^2-c_s^2)&0&0&0\\0&0&0&0\\0&0&0&0\\0&0&0&0\end{pmatrix}\right]. 
\end{equation}
In the absence of a gravitational wave, the real spacetime metric is $g_{ab}=\eta_{ab}$, where $\eta_{ab}$ has been defined in Eq. (\ref{eq:metricplane}). Therefore, the effective metric of the BEC phononic excitations on the flat spacetime metric is given by,
\begin{equation} \label{eq:phonons}
\mathfrak{g}_{ab}= \left(\frac{n^2_0\,c_s^{-1}}{\rho_0+p_0}\right) \begin{pmatrix}-c_s^2&0&0&0\\0&1&0&0\\0&0&1&0\\0&0&0&1\end{pmatrix}.
\end{equation}
Ignoring the conformal factor -which can always be done in 1D or in the case in which is time-independent- we notice that the metric is the flat Minkowski metric with the speed of light being replaced by the speed of sound $c_s$. By considering a rescaled time coordinate $\tau=(c/c_s)t$ we recover the standard Minkowski metric $ds^2=-cdt^{2}+dx^2$. This means that the phonons live on a spacetime that is Minkowski however, due to the BEC ground state properties, time flows slower and excitations propagate accordingly. As a result of this, we will show that changes in the real spacetime metric are amplified, becoming observable. 

The solutions of the Klein-Gordon equation with the metric in Eq. (\ref{eq:phonons}) describe massless excitations propagating with the speed of sound $c_s$. Therefore, the frequency of the mode $\omega_k$ is given by the dispersion relation $\omega_k=c_s\,|\bf{k}|$, where $k$ is the mode's momentum.  This linear dispersion is valid as long as $\hbar\,k<<m\,c_s$. We consider that the BEC is contained in a 1-dimensional cavity trap. Therefore, we impose close to hard-wall boundary conditions \cite{condensatebox1, condensatebox2,condensatebox3} that give rise to the spectrum,
\begin{equation}\label{eq:spectrum}
\omega_n=\frac{n\,\pi\,c_s}{L},
\end{equation}
where $L$ is the cavity length and $n\in\{1,2...\}$. The mode solutions to the Klein-Gordon equation are given by,
\begin{equation}\label{eq:minkowskimodes}
\phi_{n}=\frac{1}{\sqrt{n\,\pi}} \sin{\frac{n\pi (x-x_L)}{L}}\,e^{-i\,\omega_n\,t},
\end{equation}
where $x_L$ and $x_R$ are the positions of the left and right walls, respectively.\\
The phononic field $\Pi(t,x)$ is then quantised by associating creation and annihilation operators $a^{\dagger}_k$  and $a_k$ to the mode solutions \cite{pethicksmith},
\begin{align}
 Ê Ê\Pi(t,x)Ê Ê&= Ê\sum\limits_{k}\,\Bigl(\,\phi_{k}(t,x)\,a_{k}\,+\,\phi^{*}_{k}(t,x)\,a^{\dagger}_{k}\Bigr)\,.
 Ê Ê\label{eq:scalar field in-region}
\end{align}
 The operators $a_k$ and $a^{\dagger}_k$ obey the canonical commutation relations. In what follows we will describe the phonons in the spacetime of a gravitational wave.\\

\section {Phonon creation by a spacetime distortion}  
Consider that initially the spacetime is flat and that the phonons are in a given initial state $\sigma_0$. We are interested in computing the state of the phonons after a gravitational wave has passed by. The real spacetime metric of a gravitational wave is given by Eq. (\ref{eq:gmunu}) and thus, the effective metric for the phonons is,
\begin{equation} \label{eq:phononsgw}
\mathfrak{g}_{ab}= \left(\frac{n^2_0\,c_s^{-1}}{\rho_0+p_0}\right) \begin{pmatrix}-c_s^2&0&0&0\\0&1+h_{+}(t)&h_{\times}(t)&0\\0&h_{\times}(t)&1-h_{+}(t)&0\\0&0&0&1\end{pmatrix}.
\end{equation}
For simplicity, we considered a quasi one-dimensional BEC.  
The line element is conformal to,
\begin{equation}\label{eq:lineelement}
ds^2=-c_s^2\,dt^2+(1+h_{+}(t))\,dx^2.
\end{equation}
The effects of the spacetime distortion on the phonons can be computed using a Bogoliubov transformation \cite{birrelldavies}.
The flat field operators ${a_{k}}$ in Eq. (\ref{eq:scalar field in-region}) are transformed into,
\begin{equation}\label{bogotrans}
\hat{a}_{m}=\sum_{n} \bigl(\alpha^{*}_{mn}a_{n}+\beta^{*}_{mn}a^{\dag}_{n}\bigr)\,,
\end{equation}
where $\hat{a}^{\dagger}_k$ and $\hat{a}_k$ are creation and annihilation operators associated to the perturbed mode solutions and $\alpha_{mn}(h_+(t))$ and $\beta_{mn}(h_+(t))$ are Bogoliubov coefficients that depend on the wave's spacetime parameters. 
 In order to find the perturbed mode solutions and the Bogoliubov coefficients induced by the wave, we apply a technique developed in reference \cite{bsgates}. The technique enables the computation of Bogoliubov coefficients associated to continuous variations of spacetime by integrating discrete changes. Following this procedure, we assume that spacetime is initially flat and instantaneously undergoes a discrete spacetime perturbation $\epsilon$. The new metric will be given by,
\begin{equation}\label{eq:lineelementnot}
ds^2=-c_s^2\,dt^2+(1+\epsilon)\,dx^2.
\end{equation}
Considering the change of coordinates,
\begin{equation}\label{eq:newcoor}
t=t',\quad x=x'\,(1-\frac{\epsilon}{2}),
\end{equation} 
we find that the mode solutions to the Klein-Gordon equation in the new metric after imposing hard-wall boundary conditions at $x_L'$ and $x_R'$ are
\begin{equation}\label{eq:minkowskimodesnew}
\hat{\phi_{n}}(t'.x')= \frac{1}{\sqrt{n\,\pi}} \sin{\frac{n\pi (x'-x'_L)}{L'}}\,e^{-i\,\omega'_n\,t},
\end{equation}
where 
\begin{equation}\label{eq:lengthcoord}
L'=x_R'-x_L'=L\, ,\quad \omega'_n=\omega_n \, .
\end{equation}
We assume that the $\epsilon$-perturbation doesn't change the rigidity of the trap, that is the proper length \cite{reviewquantumgrav}:
\begin{equation}\label{eq:proplength}
L(\epsilon)=\int^{x_R(\epsilon)}_{x_L(\epsilon)}\,dx\sqrt{1+\epsilon}
\end{equation}
remains constant $L(\epsilon)=L$, up to order $\epsilon$. This means that in coordinates $(t,x)$, the boundary conditions change in the following way:
\begin{equation}\label{eq:cavitiwallscoord}
x_L(\epsilon)=x_L(1-\frac{\epsilon}{2}),\quad x_R(\epsilon)=x_R(1-\frac{\epsilon}{2}).
\end{equation}
 Thus, in coordinates $(t',x')$:
 \begin{equation}\label{eq:cavitywallscoordprime}
 x'_L=x_L;\quad x'_R=x_R
 \end{equation}
 The Bogoliubov coefficients $\alpha_{mn}(\epsilon)=(\hat{\phi}_{m},\phi_{n})$ and $\beta_{mn}(\epsilon)=-(\hat{\phi}_{n},\phi^{*}_{m})$ can be calculated using the Klein-Gordon inner product \cite{birrelldavies} between the flat and $\epsilon$-perturbed mode solutions in Eqs. (\ref{eq:minkowskimodes}) and (\ref{eq:minkowskimodesnew}). To first order in $\epsilon$ and for $x_L=0$ we find that,
\begin{eqnarray}\label{eq:bogosnot}
\beta_{mn}&=&- \frac{(-1)^{m+n}\sqrt{mn}}{2 (m+n)}\,\epsilon \,\,\,(m\neq n) \nonumber\\  \beta_{nn}&=&0\nonumber\\\alpha_{mn}&=& \frac{(-1)^{m+n}\sqrt{mn}}{2 (m-n)}\,\epsilon\, \,\,\,(m\neq n)\nonumber\\
 \alpha_{nn}&=&1
\end{eqnarray}
\footnote{The Bogoliubov identities $\sum_m |\alpha_{mn}|^2-|\beta_{mn}|^2=1$ are satisfied to first order in $\epsilon$ confirming that the total probability is conserved in the time hypersurface of integration}.  After the instantaneous perturbation the field modes undergo a period of free evolution before undergoing the next instantaneous perturbation. During free evolution the modes pick up a time-dependent phase. Following \cite{bsgates}, the Bogoliubov coefficients corresponding to the continuous perturbation produced by a gravitational wave $h_+(t)=\epsilon\,\sin{\Omega\,t}$, can be computed by approximating the perturbation $h_+(t)$ by instantaneous perturbations in $\epsilon$ followed by infinitesimal intervals of free evolution. Taking the continuous limit, we obtain the Bogoliubov coefficients corresponding to a sinusoidal perturbation,
\begin{eqnarray}\label{eq:continuousbeta}
\beta_{mn} (t)=i\,(\omega_m+\omega_n)\beta_{mn}\int^t_0 e^{-i\,(\omega_m+\omega_n)\,t'}\sin{(\Omega\,t')}\,dt'\nonumber\\
\alpha_{mn} (t)=i\,(\omega_m-\omega_n)\alpha_{mn}\int^t_0 e^{-i\,(\omega_m-\omega_n)\,t'}\sin{(\Omega\,t')dt'.}\nonumber
\end{eqnarray}
In quantum field theory, $\beta\neq0$ is associated with particle creation. Therefore, we conclude that the gravitational wave has produced phonons in the system. The number of particles produced is $n=\sum_m |\beta_{mn}|^2$ \cite{birreldavies}, therefore, proportional to the amplitude, frequency and period of the wave. 
We note that there is a particle creation resonance at
\begin{equation}\label{eq:conditio}
\Omega=\omega_m+\omega_n.
\end{equation}
At resonance, assuming that the duration of the wave $t$ is long enough,
\begin{equation}\label{eq:conditio2}
\omega_1\,t>>1, 
\end{equation}
and that $m+n=\operatorname{odd}$, the first order terms of the Bogoliubov coefficients are given by,
\begin{eqnarray}\label{eq:bogobogos}
\beta_{jk}(t)&=&\frac{\epsilon}{4}\sqrt{\frac{n}{m}}\,\omega_m\,t\,\delta_{j+k,\,m+n}\,+\mathcal{O}(\epsilon^2)\nonumber\\
\end{eqnarray} 
that is, the most relevant coefficient is $\beta_{nm}$.
The Bogoliubov coefficients we obtained coincide with those of a cavity in flat spacetime with sinusoidally varying length,
\begin{equation}\label{eq:lengthoscillation}
L(t)= L(0)(1+\epsilon\,\operatorname{sin}(\Omega\,t)),
\end{equation}
which where computed in reference \cite{koreancasimir} using an alternative method.  \\

\section{Introducing the covariance matrix formalism} 

It is convenient to use the covariance matrix formalism to describe the Bogoliubov transformation that the wave induces on the field states. Gaussian states of bosonic fields and their transformations take a very simple form in this framework. This has enabled the fast development of quantum information and quantum metrology techniques for Gaussian states and is a natural formalism for the application of quantum metrology to relativistic quantum fields \cite{rqm,rqm2}.  We start by defining the quadrature operators $X_{2n-1}=\frac{1}{\sqrt{2}}(a_{n}+a^{\dag}_{n})$ and $X_{2n}=\frac{1}{\sqrt{2}\,i}(a_{n}-a^{\dag}_{n})$, which correspond to the generalised position and momentum operators of the field, respectively.  In the covariance matrix formalism Gaussian states are completely defined by the field's first and  second moments. The first moment correspond to $\langle X_{i}\rangle$ and the second moments are encoded in the covariance matrix $\Sigma_{ij}=\langle X_{i} X_{j}+X_{j}X_{i}\rangle-2\langle X_{i}\rangle\langle X_{j}\rangle$. 
We restrict our analysis to initial Gaussian states with vanishing first moments $\langle X_{i}\rangle=0$. In this case, the state of the field after a Bogoliubov transformation is given by 
\begin{equation}\label{eq:transformedcm}
\sigma_{\epsilon}=S_\epsilon\sigma_0 S_\epsilon^{T},
\end{equation}
where $\sigma_0$ encodes the initial state of the field and the Bogoliubov transformation in Eq.~(\ref{eq:bogobogos}) is encoded in the symplectic matrix\be\label{Bogosymplectic}
S=\left(
  \begin{array}{cccc}
    \mathcal{M}_{11} & \mathcal{M}_{12} & \mathcal{M}_{13} & \cdots \\
    \mathcal{M}_{21} & \mathcal{M}_{22} & \mathcal{M}_{23} & \cdots \\
    \mathcal{M}_{31} & \mathcal{M}_{32} & \mathcal{M}_{33} & \cdots \\
    \vdots & \vdots & \vdots & \ddots
  \end{array}
\right)\,,
\ee
where the $2\times2$ matrices $\mathcal{M}_{mn}$ are given by
\be\label{Mmatrices}
\mathcal{M}_{mn}=\left(
                   \begin{array}{cc}
                     \Re(\alpha_{mn}-\beta_{mn}) & \Im(\alpha_{mn}+\beta_{mn}) \\
                     -\Im(\alpha_{mn}-\beta_{mn}) & \Re(\alpha_{mn}+\beta_{mn})
                   \end{array}
                 \right)\,.
\ee
$\Re$ and $\Im$ denote the real and imaginary parts, respectively. The Bogoliubov coefficients and thus, the final state of the field, depend on the amplitude of the gravitational wave $\epsilon$. Our main aim is now to determine under what circumstances these changes are observable and moreover, use specialised techniques in relativistic quantum metrology to estimate the amplitude of the gravitational wave by making measurements on the BEC phononic field.\\

\section{Relativistic Quantum Metrology: estimating the amplitude of the wave.} Quantum metrology provides techniques to estimate parameters associated to the transformation of a quantum state. The formalism provides strategies, that include finding both, initial states and measurements basis, that enable one to estimate the parameter with optimal precision. Recently a formalism to estimate parameters in relativistic quantum fields has been developed in the covariance matrix formalism \cite{rqm,rqm2}. In this section we apply these techniques to provide a bound on the optimal precision that can be achieved when estimating the wave amplitude $\epsilon$ through measurements on $\sigma_\epsilon$. The quantum Cramer-Rao theorem states that the error in the estimating the parameter $\epsilon$ is bounded by~\cite{BraunsteinCaves1994}, 
\be\label{Cramer-Rao}
\langle (\Delta \hat{\epsilon})^{2}\rangle\geq \frac{1}{MH_\epsilon}.
\ee
where $H_{\epsilon}$ is the Quantum Fisher Information (QFI) and $M$ the number of probes. The QFI can be computed using the Uhlmann fidelity~$\mathcal{F}$ between the state $\sigma_{\epsilon}$ and a state with an infinitesimal increment in the parameter, i.e. $\sigma_{\epsilon+d\epsilon}$,
 \begin{equation}
 H_\epsilon=\frac{8\big(1-\sqrt{\mathcal{F}(\sigma_{\epsilon},\sigma_{\epsilon+d\epsilon})}\big)}{d\epsilon^{2}}.\label{quantumfishinfo}
\end{equation}
The precision in the estimation of the parameter will be increased when $\sigma_{\epsilon}$ and $\sigma_{\epsilon+d\epsilon}$ are more distinguishable. 
Now let $\sigma_{\epsilon}$ be  a two-mode Gaussian state with zero initial first moments. The Fidelity between them is given by
\cite{MarianMarian}
\begin{eqnarray}\label{MarianMarian}
    \mathcal{F}(\sigma_{\epsilon},\sigma_{\epsilon+d\epsilon}),=\frac{1}{\sqrt{\Lambda}+\sqrt{\Gamma}-\sqrt{(\sqrt{\Lambda}+\sqrt{\Gamma})^2-\Delta}},\label{two:mode:fidelity}
\end{eqnarray}
where
\begin{eqnarray}
\Gamma =&\,\frac{1}{16}\text{det}(i\mathbf{\Omega}\sigma_{\epsilon} i\mathbf{\Omega}\sigma_{\epsilon+d\epsilon}+\mathbbm{1})\nonumber\\
\Lambda =&\,\frac{1}{16}\text{det}(i\mathbf{\Omega}\sigma_{\epsilon}+\mathbbm{1})\text{det}(i\mathbf{\Omega}\sigma_{\epsilon+d\epsilon}+\mathbbm{1})\nonumber\\
\Delta =&\,\frac{1}{16}\text{det}(\sigma_{\epsilon}+\sigma_{\epsilon+d\epsilon})\nonumber\\
\end{eqnarray}
where $\mathbbm{1}$ is the identity matrix and the symplectic form $\Omega$ is given by $\Omega=\bigoplus_{k=1}^{n}\Omega_k$, $\Omega_k=-i\sigma_y$ and $\sigma_y$ is one of the Pauli matrices.

In reference \cite{rqm,rqm2} analytical formulas are provided for the computation of the QFI in terms of Bogoliubov coefficients that admit a perturbative expansion in terms of the parameter to be estimated. 
We consider the state of the field to be a two-mode squeezed state for two modes $n$ and $m$ with squeezing parameter $r$ -the rest of the modes are initially in the vacuum state. In the covariance matrix formalism, the reduced state of a particular set of modes is obtained by simply deleting the corresponding rows and columns of the covariance matrix.  
The QFI for this particular initial state was computed in \cite{rqm2} in terms of general Bogoliubov coefficients, 
\begin{eqnarray}\label{H3}
H&=&\epsilon^{-2}\Re\bigg[4 \cosh r(f^{n}_{\alpha}+f^{n}_{\beta}+f^{m}_{\alpha}+f^{m}_{\beta})
+ 4\cosh^2r(2|\beta_{nm}(t)|^2-f^{n}_{\alpha}+f^{n}_{\beta}\nonumber\\&-&f^{m}_{\alpha}+f^{m}_{\beta})
+4\sinh^2r(f^{n}_{\alpha}-f^{n}_{\beta}+f^{m}_{\alpha}-f^{m}_{\beta}-2\beta_{nm}(t)^{2}+2\alpha_{nm}(t)^{2})
\nonumber\\&+&4\sinh r\, \Re[\mathcal{G}^{\alpha \beta}_{nm}+\mathcal{G}^{\alpha \beta}_{nm}]-4\cosh^4 r |\beta_{nm}(t)|^2
-\frac{1}{2}\sinh^22r(2|\alpha_{nm}(t)|^2\nonumber\\&-&3|\beta_{nm}(t)|^2
-\beta_{nm}(t)^{2}\bigg].\nonumber\\
\end{eqnarray} 
where
\begin{eqnarray}
f_{\alpha}^{i}&=&\,\frac{1}{2}\sum_{j\neq n,m}|\alpha_{ji}|^2\nonumber\\
f_{\beta}^{i}&=&\, \frac{1}{2}\sum_{j\neq n,m}|\beta_{ji}|^2\nonumber\\
\mathcal{G}^{\alpha\beta}_{ij}&=&\,\sum_{k\neq n,m} \alpha_{ki}}{\beta_{kj}^{*}\nonumber\\.
\end{eqnarray} 
Substituting Eq. (\ref{eq:bogobogos}), we obtain,
\begin{eqnarray}\label{eq:fish}
H_\epsilon=\frac{n}{4m}\omega^2_m\,t^2\,(8-4\cosh^4(r)+2\sinh^2(2r)).
\end{eqnarray}\\

\section {Is the effect observable?} 
\begin{figure}[t!]
\includegraphics[width=0.8\linewidth]{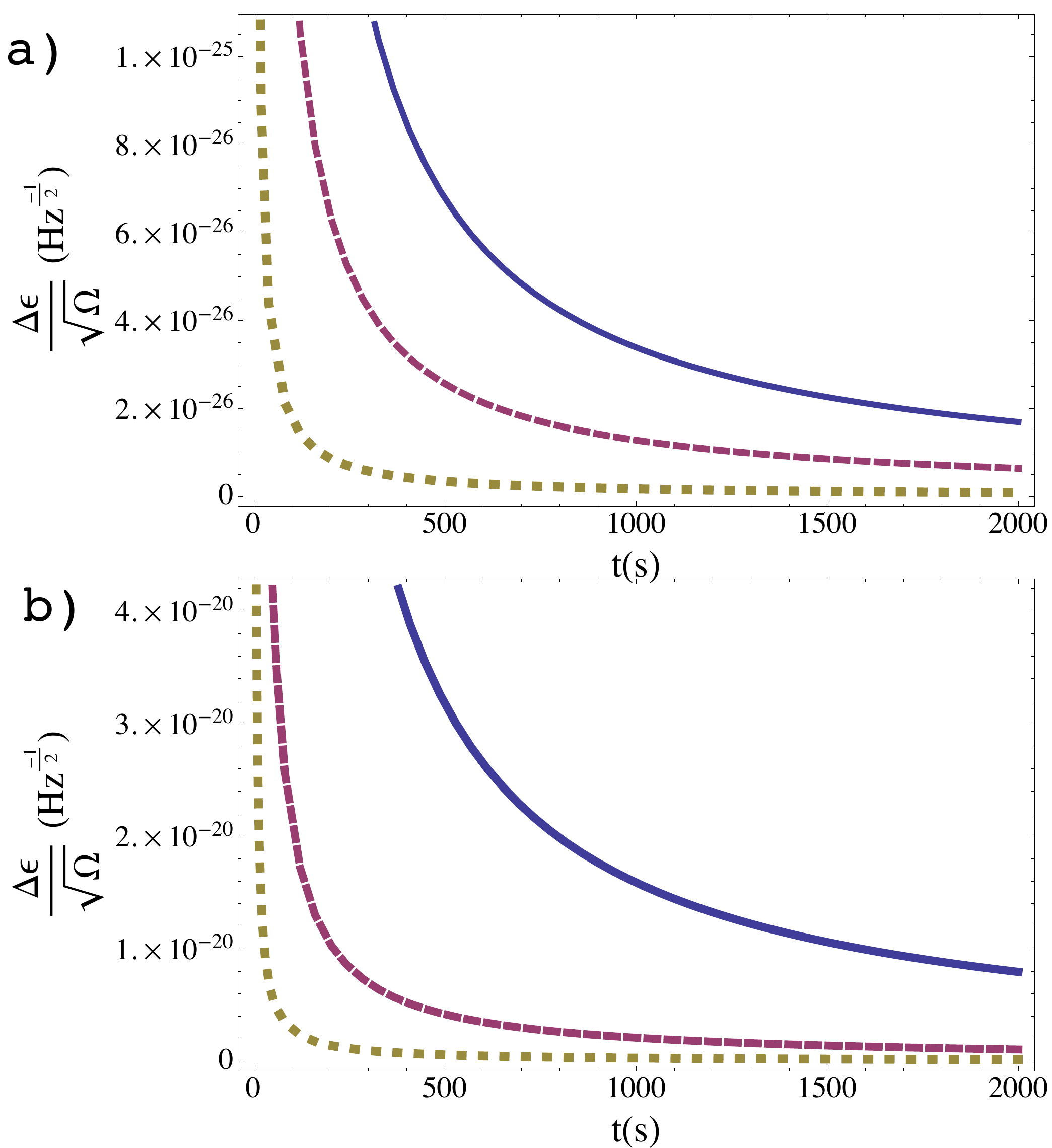}
\caption{Optimal bound of the strain sensitivity $\frac{\Delta\epsilon}{\Omega}$ versus duration of the gravitational wave for $L=1\, \operatorname{\mu m}$, $c_s=10\,\operatorname{mm/s}$, $M=10^{14}$. a) $r=10$ and mode numbers $m=1$, $n=2$ (blue, solid), $m=1$, $n=6$ (red,dashed), $m=10$, $n=11$ (yellow,dotted). b)$m=10$, $n=11$ and squeezing parameter $r=2 $ (blue, solid), $r=3$ (red,dashed) and $r=4$ (yellow, dotted).}\label{fig:fig2}
\end{figure}

The fact that gravitational waves can generate photons in fields confined in certain regions of spacetime was pointed out in \cite{dograv,dograv2}. However, the effect is negligible in optical cavities since gravitational waves typically have frequencies several orders of magnitude below the optical regime. The condition for a particle creation resonance given by Eq. (\ref{eq:conditio}) in the case of an optical cavity, would require waves in the PHz regime. Fortunately, the situation is very different for the phononic excitations of a BEC. The slow propagation of the excitations acts as an amplification effect making particle creation observable at regimes that can be in principle reached in the experiment. Considering a cavity with length $L=1\,\operatorname{\mu m}$ and typical values for the speed of sound in such systems $c_s=10\,\operatorname{mm/s}$ results in a fundamental frequency of $\omega_1=2\pi\times 5000\,Hz$, which contrasts with the typical fundamental frequencies in optical cavities, which are in the PHz regime. Experimental timescales for the duration of the wave between 0 and 2000 seconds, would allow to detect persistent sources such as the stochastic cosmic gravitational wave background \cite{ligo1}, as well as short and long-lived \cite{ligo2} gravitational waves transients such as gamma-ray bursts. Typical phononic states on a BEC last approximately a few seconds, however, for very low densities the state's lifetime can be between 100-1000 seconds. We consider values of the initial squeezing parameter between $r=2$ and $r=10$ that seem in principle achievable in time-dependent potential traps \cite{serafinisqueezing}. Each pair of mode numbers in turn provides a different value of the resonant frequency of the gravitational wave. This fact can be exploited to detect waves in different frequencies.  Using these parameters, we plot in Figs. (\ref{fig:fig2}a) and (\ref{fig:fig2}b) our bound to a typical figure of merit that is used in the literature of gravitational wave astronomy, i. e. strain sensitivity  \cite{gravwavesdetectors}. This figure of merit is given by $\Delta\epsilon/\sqrt{\Omega}$. 
The range of frequencies considered are between $10^4-10^5\,\operatorname{Hz}$, which partially overlaps with the high frequency band of LIGO and also enters into the high-frequency realm \cite{hfgw}. The condition $\hbar\,k<<m\,c_s$ holds for these values of the frequencies and, for instance, the mass of ${}^{87} Rb$. Note that thermal noise is negligible in the considered range of frequencies at experimentally achievable temperatures. For instance, at $10 \operatorname{nK}$ the average number of phonons of $15 \operatorname{KHz}$ is $5\times 10^{-32}$. Temperatures as low as $0.5 \operatorname{nK}$ -at which the average number of phonons in that frequency is $10^{-625}$- have been achieved in the laboratory \cite{sciencewr}.

The number $M$ is assumed to be $M=10^{14}$. This number is assumed in non-relativistic proposals to detect gravitational waves with BECs in space \cite{sagas, sagas2}. In these schemes the wave function of the BEC is split and recombined using laser pulses, giving rise to a phase shift of $\phi=k\, a\, T^2$, where $k$ is the wave number of the atomic hyperfine transition, $a$ is the average acceleration and $T$ the interrogation time between pulses. The readout of the phase is then performed by fluorescence imaging of the atoms. The optimal sensitivity in the measurement of acceleration as provided by the QFI is given by \cite{sagas2}:
$\delta a =1/(\sqrt{N}\,  k\, T^2 )$.
After one cycle, N is given by the number of detected atoms. After several cycles, the number of atoms is multiplied by the repetition rate and the integration time. Considering \cite{sagas2} a number of atoms of $10^6$ -which already assumes a signal-to-noise ratio with respect to the total number of atoms in the BEC-, a repetition rate of $5 \operatorname{Hz}$ and integration time of 1 year, a number $N= 10^{14}$ is obtained. The acceleration can in turn be related with the amplitude of a gravitational wave. Note that integration times of several years are achievable by stable large-atom BEC machines \cite{becmachines, becmachines2}. 

Let us discuss several alternative measurement protocols in our case. First, the readout of the changes in the covariance matrix might be performed in a manner similar to the experiment in \cite{casimirwestbrook}, where upon releasing the condensate trapping potential, the state of each phonon is mapped into the state of an atom with the same momentum and the velocities are measured by a position-sensitive single-atom detector. Then again number of atoms, repetition rate and integration time are the relevant parameters to determine the total huber of probes, and similar numbers as above can be considered. Note however that even with only one cycle of measurements ($M=10^6$), the sensitivity of our device is comparable to the state-of -the-art in laser interferometry, namely $7\times 10^{-23}\operatorname{Hz}^{-1/2}$. 

An alternative non-destructive method of readout consists in using atomic quantum dots or optical lattices coupled to the condensate to probe the field state \cite{nuestrotermo}.  The interaction between each dot and the condensate can be modulated through Feshbach resonances in the sub-ms regime \cite{feshbachlucia} and a number of 1500 dots can be considered \cite{nuestrotermo}. This results in the possibility of making $10^6$ measurements in $1 \operatorname{s}$, giving rise to the same number of measurements after one year of integration time. Another method to measure the covariance matrix of a pair of phononic modes through non-destructive measurements has been recently introduced in \cite{carusottofinazzi}. Note also that in \cite{megamind} the authors report on experimental measurements of quantum fluctuations of the number of phonons in a particular mode, by using in-situ techniques. 
As can be seen in Figs. (2a) and (2b), we find that the strain sensitivity can be as low as $10^{-27}\operatorname{Hz}^{-1/2}$ for the optimal parameters. This is well below the threshold for the detection of gravitational waves. State-of-the-art in LIGO is $7\times 10^{-23}\operatorname{Hz}^{-1/2}$, which in our case can be achieved with $10^6$ measurements.

Our intention in this paper is to suggest that gravitational waves can be detected through the excitation of the phononic modes of a BEC and to show, using realistic parameters, that the effects are in principle observable.  A detailed experimental proposal will be provided elsewhere. However, let us briefly discuss which challenges would be necessary to overcome, such as depletion and rigidity of the trap. 

The experiment would require very low temperatures in order to safely neglect both thermal and quantum depletion. However, as discussed above, temperatures as low as 0.5 nK have been achieved in the laboratory, where the effects mentioned are negligible. Assuming that the system is cold enough  and conveniently isolated from other sources of vibration, the main challenge to address will be then to achieve the required rigidity of the trap. Signals produced by random motion of the boundaries, due to for instance, beam-pointing laser noise, should be clearly distinguishable from the gravitational wave signal, whose features depend critically on the relevant frequencies and on its particular way of transforming the spacetime coordinates (Eq. \ref{eq:cavitiwallscoord}). In particular, beam-pointing fluctuations in optical traps are completely negligible in the kHz regime considered in this work \cite{pra97, OpticsExpress12}, so they will not generate a competing signal. The only effect of the beam-pointing fluctuations is an additional heating, designed to be negligible in modern optical traps \cite{dukegroup}. An in-depth analysis of possible sources of non-rigidity depends strongly on the particular type of trap and the details of the experimental setup. This analysis will be presented elsewhere.

\section{Conclusions} 
We have proposed a method for gravitational-wave astronomy based on a relativistic quantum field theoretical approach. We have shown that spacetime distortions produce phonons in a Bose-Einstein condensate and we have suggested to use this effect to detect gravitational waves.  The spacetime wave gives rise to particle creation resonances when the BEC is in a box-like potential, similar to the Dynamical Casimir Effect.  Particle creation through the motion of boundary conditions has been extensively analysed in the context of the Dynamical Casimir Effect \cite{moore, casimirwilson}. However, to the best of our knowledge, this is the first time it is shown that the real dynamics of spacetime can generate phonons in a BEC. Since the final state of the phonons depends on the amplitude of the wave, the amplitude can be estimated. We calculated a bound on the optimal precision that can be achieved with this method and studied the regime of experimental parameters in which the sensitivity is low enough to detect the spacetime ripple.  In the best scenario, the predicted strain sensitivity is several orders of magnitude beyond the performance of highly sophisticated programmes for gravitational wave detection such as LIGO. The experiment clearly presents challenges, however, introducing new methods for the detection of spacetime effects might not only lead to the detection of gravity waves but also deepen our understanding of the overlap of quantum theory and relativity. The techniques presented in this paper can be extended to explore the effects of other spacetimes including expanding metrics.  We hope our scheme for relativistic quantum metrology will not only play a role in the upcoming era of gravitational wave astronomy but also show how the interplay between quantum and relativistic effects can give rise to a new generation of quantum technologies. 

\section*{Acknowledgements} We thank Tupac Bravo, Jorma Louko and Chris Westbrook for useful discussions and comments. C. ~S, M.~A. and I.~F. acknowledge support from EPSRC (CAF Grant No.~EP/G00496X/2 to I.~F.).

\end{document}